\documentclass[aps,twocolumn,showpacs]{revtex4}
\usepackage{graphicx}
\begin{document}
\title{Long Range Interactions in Quantum Many Body
Problem in One-Dimension:Ground State }
\author{Saugata Ghosh}\email{saugata@prl.ernet.in}
\affiliation{Physical Research Laboratory, Navrangpura, Ahmedabad -380009
India.}
\date{\today}
\begin{abstract} We investigate the ground state properties
 of  a family of $N$-body systems 
in 1-dimension, trapped
in a polynomial potential and 
 having long range $2$-body interaction
in addition to the inverse square potential
studied in the Calogero-Sutherland model (CSM). We show  
that for such a Hamiltonian, the
ground state energy is similar to
that of free fermions in a harmonic well
with a displacement that depends on the
number of particles and depth of the well.
We obtain the ground state wave function and
using random matrix results, study 
the particle density and pair correlation function (PCF). We
observe that the particles are arranged in bands.
Due to the presence of long range interaction, the
PCF shows a departure from  the CSM.
\end{abstract}
\pacs{03.75.Kk,05.30.Jp,03.65.Ge}
\maketitle{}

Theoretical understanding of the ground state properties
of complex many body systems have received considerable
attention in recent years
\cite{Sutherland,Calogero,Panigrahi2,Panigrahi3,Papenbrock1,Sogo,Lloyd,Papenbrock2,Forrester,Panigrahi1,Vignolo,Beenakker,Ketterle}. 
In this context, we study rigorously a wide
class of $1$-dimensional $N$-body systems having different densities,
nature and strength of a $2$-body interaction. We obtain
the ground state properties of these 
systems having a $2$-body potential
$V_{2}=g/r_{12}^{2}+\Phi_{l}$, where $r_{12}$
is the interparticle spacing and $\Phi_{l}$
contains the long range $2$-body interaction.
The system is trapped in a 
polynomial potential. This may be relevant in understanding
the various aspects of the Bose-Einstein-Condensates, where a wide
variety of potentials under a controlled environment is possible.

We  derive the  ground state eigenvalues and eigenfunctions
for such systems and extract several interesting properties 
by identifying the square of the wave function
with the joint probability distribution (j.p.d.)
of eigenvalues of non-Gaussian ensembles of
random matrices. Using the polynomial method
developed by Ghosh and Pandey in  the context
of random matrix theory (RMT) \cite{pg,gp}, we
observe band structure \cite{gpuri} in the particle density,
which in turn corresponds to the density of zeros
of the corresponding polynomials \cite{Lloyd,pg,gp}.
For a given value of the interaction strength, we
study the PCF for different
interparticle spacings. Due to the presence
of the long range interaction, we observe a deviation
from the CSM \cite{Sutherland,Calogero}.

We shall consider the ground state  of a system  of $N$
particles satisfying the Schr\"{o}dinger equation
\begin{eqnarray}
\label{Hamiltonian}
\left[-\sum_{i=1}^{N}\frac{{\partial}^{2}}{\partial {x_{i}}^2}
+\prod_{i<j}\frac{g}{{(x_i-x_j)}^{2}}
+\Phi_{l}(x_i,x_j)+V_{1}(x_i)-E_{n}\right]{\psi}_n 
=0,
\end{eqnarray}
where terms corresponding to long range interaction
derive from the relation
\begin{equation}
\Phi_{l}(x_i,x_j)={}
h\sum_{i\ne j}\frac{P(x_i)}{(x_i-x_j)},
\end{equation}
$P(x)$  being any analytic function 
having a power expansion in $x$, while $g$ and $h$ are the
interaction strengths. The system is trapped
in a potential 
\begin{equation}
V_{1}(x_{i})=\sum_{i=1}^{N}[P^{2}(x_{i})-P^{\prime}(x_{i})].
\end{equation}
In this paper, we  consider the case where $P(x)$ is a 
polynomial of order $2m+1$ and is represented by
\begin{equation}
\label{pot}
P(x)=\gamma\sum_{k=0}^{m}a_{2k+1}x^{2k+1}.
\end{equation}
For convenience, we  take $a_{2m+1}=1$.
The parameter $\gamma$ will determine the depth of the well.
The case where $m=0$ corresponds to the CSM \cite{Sutherland},
where the $2$-body interaction is purely of the inverse square type.
For $m>0$, we will encounter the long range interaction. 

For such a Hamiltonian, the ground state wave function
can be written as

\begin{equation}
\label{psii}
\psi_{0}\equiv\phi\varphi=\prod_{i<j}{|x_i-x_j|}^{\lambda}
\exp[-\sum_{i}\int_{0}^{x_i}P(t_i)dt_i].
\end{equation}
It should be noted that if the 
particles are  confined to a configuration space 
$x_{1}>x_{2}\ldots >x_{N}$,
the $g/r^{2}$ interaction in one dimension does not allow them
to cross, thereby respecting the ordering. This is valid
provided the potential is not too attractive, which leads to
the restriction $g\ge -1/2$ \cite{Sutherland,Landau}. For such
an ordering, the modulus becomes irrelevant and hence depending
on the value of $\lambda$ (odd or even) the system
is  bosonic or fermionic in nature.
Also, for such a  configuration,
$\psi_{0}$ is nodeless (apart from the trivial ones at the
points of coincidence) and hence corresponds
to the ground state of the system.
Here
\begin{equation}
\phi=\prod_{i<j}{|x_i-x_j|}^{\lambda},
\end{equation}
\begin{equation}
\label{varphi}
\varphi=\exp[-\sum_{i}\int_{0}^{x_i}P(t_i)dt_i],
\end{equation}
\begin{equation}
\label{g}
(\lambda^2-\lambda)=g/2,
\end{equation}
and 
\begin{equation}
\label{hh}
2\lambda=-h.
\end{equation}
The ground state energy  is given by
\begin{equation}
\label{eigen}
E_0=-\gamma a_{1}N(1+\lambda(N-1)).
\end{equation}
For $a_{1}>0$, this is the same as that of free fermions in a harmonic well
$\gamma a_{1}$, displaced by a factor $\gamma\lambda Na_{1}(N-1)$.

First we prove that for $P(x_{i})$ given by Eq.(\ref{pot}),
Eq.(\ref{psii}) and Eq.(\ref{eigen}) are respectively the ground state
eigenfunction and eigenvalues. We study two cases,
where we calculate $\psi_{0}$ and $E_{0}$ and recognize
$\psi_{0}^{2}$  as the j.p.d.
of eigenvalues for the non-Gaussian ensembles of random matrices.
Using the polynomial method, we study the particle
density  and the effect of long range interaction on the
PCF. PCF shows a deviation
from the CSM, which thereby suggests a departure
from the ``universal'' result of RMT.

The kinetic  term of the Hamiltonian  acting on $\psi_{0}$ gives
\begin{eqnarray}
\nonumber
-\sum \frac{{\partial}^{2}(\phi\varphi)}{\partial {x_{i}}^2} =
-\sum \varphi\frac{{\partial}^{2}(\phi)}{\partial {x_{i}}^2} -
&2&\sum \frac{{\partial}(\varphi)}{\partial {x_{i}}}
.\frac{{\partial}(\phi)}{\partial {x_{i}}}\\
&-&\sum \phi\frac{{\partial}^{2}(\varphi)}{\partial {x_{i}}^2} .
\end{eqnarray}
Substituting  $\psi_{0}$ from  Eq. (\ref {psii}), the first term gives 
\begin{equation}
\sum \varphi\frac{{\partial}^{2}(\phi)}{\partial {x_{i}}^2}=
[2(\lambda^2-\lambda)\sum_{i<j}\frac{1}{{(x_i-x_j)}^{2}}]\phi\varphi.
\end{equation}
The second term gives 
\begin{equation}
\label{second}
-2\sum \frac{{\partial}(\varphi)}{\partial {x_{i}}}
.\frac{{\partial}(\phi)}{\partial {x_{i}}}
=2\lambda\phi\varphi\sum_{i\ne j}\frac{P(x_i)}{(x_i-x_j)},
\end{equation}
while it is easy to see that the last term
\begin{equation}
\label{third}
-\sum \phi\frac{{\partial}^{2}(\varphi)}{\partial {x_{i}}^2} 
= \sum_{i}[P^{\prime}(x_i)- P^{2}(x_i)]\phi\varphi.
\end{equation}
Thus we see that our choice of $\psi_{0}$  diagonalises
the Hamiltonian  (\ref {Hamiltonian}).
Replacing $P(x_i)$ from Eq.(\ref {pot})
in Eqs.(\ref {second}) and (\ref {third}),
one can easily obtain
the ground state energy  for the cases where $m \ge 0$.

We will consider the two cases, corresponding to
\begin{equation}
\label{pot1}
P(x_i)= \gamma(x_{i}^3-a_{1} x_{i}),\hspace{1cm}\gamma>0,
\end{equation}
and
\begin{equation}
\label{pot2}
P(x_i)=\gamma(x_{i}^5-a_{3}x_{i}^3+a_{1}x_{i})
\hspace{1cm}\gamma>0, a_{3}<0.
\end{equation}
They will  not only illustrate  the formation of 
multiple bands in the particle density for appropriate
values of the parameter $a_{1}$, but also show the
effect of long range interaction in determining the
PCF.

The Hamiltonian corresponding to Eq.(\ref{pot1}) is,
\begin{eqnarray}
\label{Ham1}
\nonumber
H = -\sum \frac{{\partial}^{2}}{\partial {x_{i}}^2}
+{\prod_{i<j}}\frac{g}{{(x_i-x_j)}^{2}}
-\frac{h\gamma}{2}\sum_{i<j}{(x_{i}-x_{j})}^{2}\\
+{\gamma}^{2}\sum_{k=1}^{3}b_{2k}x_{i}^{2k}.
\end{eqnarray}
For $1>>\gamma>0$, the system behaves
like an anisotropic oscillator, kept in a weak polynomial well. For
such systems, the motion is bounded for all physically possible energies.
In the thermodynamic limit, i.e. for $N\rightarrow\infty$ and
$\gamma\rightarrow 0$, we will have a finite particle density, whose
shape will depend on the well it is kept in.
Finally, repeating the steps outlined earlier, one can
find that a choice of $\psi_{0}$
\begin{equation}
\psi_{0}\equiv\phi\varphi=\prod_{i<j}{|x_i-x_j|}^{\lambda}
\exp\left[-\gamma\left(\sum_{i}\frac{x_{i}^4}{4}
-a_{1}\frac{x_{i}^2}{2}\right)\right],\hspace{0.3cm}\gamma>0
\end{equation}
satisfies the Schr\"{o}dinger equation, with $g$ and $h$ given by
Eqs. (\ref {g}) and (\ref {hh}) respectively, and
\begin{equation}
b_{6}=1, b_{4}=-2a_{1}, b_{2}= {a_{1}}^{2}-(6\lambda+3)/\gamma.
\end{equation}
For $\gamma {a_{1}}^{2}=6\lambda+3$ and $\gamma\rightarrow 0$, we can
obtain the ground state wave function and eigenvalue for the central
potential $V(r_{12})=Ar_{12}^{2}+B/r_{12}^{2}$ and compare the result
for $N=2$ with that obtained in Ref.\cite{Landau}.
For such a choice,
we find $E_{0}=2(1+\lambda)\sqrt{(3\gamma(2\lambda+1))}$ as compared
to $E_{0}=\sqrt{\lambda\gamma}(2\lambda+1)$ obtained in \cite{Landau}.
The deviation is due to the effect of the potential well. For $N$-particles,
the ground state energy is given by Eq.(\ref{eigen})
 with a negative sign introduced
due to the negative coefficient  $a_{1}$. 
The Hamiltonian corresponding to Eq.(\ref{pot2})  can be written as,

\begin{eqnarray}
\label{Ham2}
\nonumber
H&=&-\sum \frac{{\partial}^{2}}{\partial {x_{i}}^2}
+{\prod_{i<j}}\frac{g}{{(x_i-x_j)}^{2}}
+\alpha_{1}\sum_{i<j}{(x_{i}-x_{j})}^{4}\\
&+&\alpha_{2}\sum_{i<j}{(x_{i}-x_{j})}^{2}
+\alpha_{3}\sum_{i<j}{(x_{i}+x_{j})}^{4}
+\gamma^{2}\sum_{k=1}^{5}b_{2k}x_{i}^{2k}.
\end{eqnarray}
Here, the second, third and fourth terms have the same effect of making
the particles remain in a bound state, with the parameter $\gamma$
and $a_{3}$ controlling respectively the range and depth of the potential
.

Repeating the steps outlined earlier, one can
find that a choice of $\psi_{0}$
\begin{eqnarray}
\psi_{0} =\prod_{i<j}{|x_{i}-x_{j}|}^{\lambda}
\exp\left[-\gamma\left(\sum_{i}\frac{x_{i}^6}{6}
-a_{3}\sum_{i}\frac{x_{i}^4}{4}
+a_{1}\frac{x_{i}^2}{2}\right)\right],\hspace{0.25cm}\gamma>0,
\end{eqnarray}
with $\gamma>0$, $a_{3}<0$ satisfies the Schr\"{o}dinger equation, with
$h$ given by Eq.(\ref{hh}), and
\begin{eqnarray}
\nonumber
&&\alpha_{1}=\lambda\gamma/12,\alpha_{2}=-\lambda\gamma a_{3},
\alpha_{3}=-5\lambda\gamma/12,\\
\nonumber
&& b_{10}=1,b_{8}=-2a_{3},b_{6}=(a_{3}^{2}+2a_{1}),\\
&& b_{4}=-(2a_{1}a_{3}+(5+10\lambda/3)/\gamma),
b_{2}=a_{1}^{2}+a_{3}(6\lambda+3)/\gamma.
\end{eqnarray}
The ground state energy is given by Eq.(\ref {eigen}).
As in the previous case,  the parameter $a_{1}$ will not only
determine the  position of the ground state
but will also be crucial in determining  the band structure.

It is at this point that, we will write $\psi_{0}$ in terms
of the variables

\begin{equation}
y_{i}={\left(\frac{1}{\lambda}\right)}^{1/(2m+2)}x_{i}.
\end{equation}
Then $\psi_{0}$ can be written as
\begin{equation}
\psi_{0}=C^{1/2}\prod_{i<j}{|y_{i}-y_{j}|}^{\lambda}
\exp[-\lambda\sum_{i} \int_{0}^{y_{i}}P_{1}(t_{i})dt_{i}],
\end{equation}
where 
\begin{equation}
P_{1}(y_{i})=\gamma\sum_{k=0}^{m}c_{2k+1}y_{i}^{2k+1},
\end{equation}
is  a monic polynomial of order $2m+1$, whose coefficients are related
to those of $P(x_i)$ by
\begin{equation}
c_{2k+1}={\left(\frac{1}{\lambda}\right)}^{(m-k)/(m+1)}a_{2k+1}.
\end{equation}
Then ${\psi_{0}}^{2}$ is given by
\begin{equation}
{\psi_{0}}^{2}=C\prod_{i<j}{|y_{i}-y_{j}|}^{\beta}
\exp[-\beta\sum_{i}\int_{0}^{y_{i}}P_{1}(t_{i})dt_{i}].
\end{equation}
with $\beta=2\lambda$,  and $C$ being the  normalization constant. 
Now one may  interpret ${\psi_{0}}^{2}$ to be identical
with the j.p.d. of non-Gaussian ensembles of random
matrices \cite{gp}. We define
the $n$-particle correlation function 

\begin{equation}
R_{n}(y_{1}, y_{2},\ldots, y_{n})=\int\ldots\int dy_{n+1}\ldots dy_{N}{\psi_{0}}^{2}.
\end{equation}
as the probability of finding $n$-particles in the intervals
$y_{i}$ and $y_{i}+\Delta y_{i}$, irrespective of the position
of the other particles. 
$n=1$ and $2$ correspond to  the particle density and PCF respectively.
It is shown in Ref. \cite{gp} that $R_{1}(x)$ corresponds to the density 
of zeros of the polynomial having weight function given by (\ref{varphi}).
It has been shown by Dyson in the context of random matrices
that for $\beta=1$, $2$ and $4$, 
 $R_{n}$ can be written in terms of orthogonal
and skew-orthogonal polynomials.
For $\beta=2$, the PCF  can be written as 
\begin{equation}
\label{R2}
R_{2}(y_{1},y_{2})=\sum_{\mu=0}^{N-1}{h_{\mu}}^{-1}[q_{\mu}(y_{1})q_{\mu}(y_{2})]
\exp[-2\int _{0}^{y_{2}}P_{1}(t)dt]
\end{equation}
where $q_{\mu}(y)$ are orthogonal polynomials corresponding to the
normalization condition
\begin{equation}
\int_{-\infty}^{\infty}q_{\mu}(y)q_{\nu}(y)
\exp[-2\int _{0}^{y}P_{1}(t)dt]dy=h_{\mu}\delta_{\mu\nu}.
\end{equation}

\begin{figure}
\scalebox{0.4}{\includegraphics{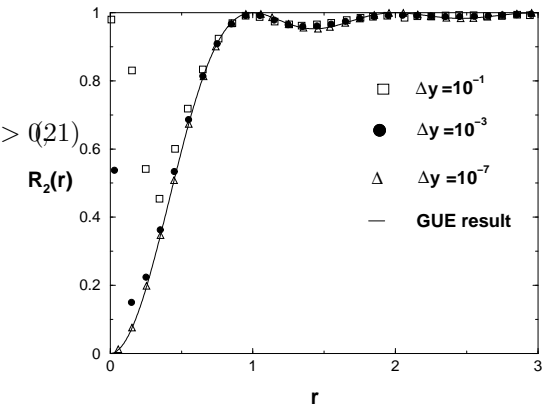}}
\caption{shows the smoothed PCF \cite{gp} for the Hamiltonian
(\ref{Ham1}), with $\beta=2$, $N=50$, $a_{1}=60$ and $r=\Delta y.R_{1}(y)$.
The curves correspond to different values of $\Delta y$.
The solid line corresponds to CSM (GUE) result, with $\Delta y=10^{-2}$.}
\end{figure}

Now, we come back to the
Hamiltonian discussed in Eq.(\ref {Ham1}).
For $a_{1}>0$, the ground state energy is
positive. For
$a_{1}\equiv a_{c}=\sqrt{2N/\gamma}$, we define 
 $E_{0c}=\gamma a_{c}N(1+\lambda(N-1))$.
For $E_{0}> E_{0c}$,
we observe  the formation of  two bands
in particle density. Rescaling the result obtained by Pandey
\cite{gpuri}, we get

\begin{eqnarray}
\label{pandey1}
\pi R_{1}(y) &=& \gamma|y|\sqrt{2N/\gamma-{( y^2 -a_{1})}^{2}}.
\end{eqnarray}

For $-E_{0c}<E_{0}<0$, a passage through the barrier is possible
resulting in a splitting of each of these levels into two
neighboring ones, corresponding to the state in which the
particles move simultaneously through
both the barriers. This corresponds
to a single band case with a dip around the origin. Finally, for
$E_{0}<-E_{0c}$,
we get a single band with a maximum at $y=0$.
This is given by

\begin{eqnarray}
\nonumber
\pi R_{1}(y) &=& \gamma
\left[\frac{1}{3\gamma}\{\sqrt{{a_{1}}^{2}+6N/\gamma}-2a_{1}\}+y^2\right]\\
&&\times {\left[\frac{2}{3\gamma}\{\sqrt{{a_{1}}^{2}+6N/\gamma}+a_{1}\}
-y^2\right]}^{1/2},
\end{eqnarray}
for $a_{1}\le \sqrt{2N/\gamma}$  for the single band case.
We observe that for finite $R_{1}(y)$, letting $N\rightarrow\infty$
implies that $\gamma\rightarrow 0$ as $N^{-1}$.
It should be noted that in the thermodynamic limit, the particle density, after
proper scaling, is
independent of $\beta$ and hence applicable for both bosons and fermions.
However the calculation of the PCF,
essential for the study of thermodynamic properties,
is restricted to $\beta=1$, $2$ and $4$. 
It has been proved in Ref.\cite{pg,gp} 
that for interparticle spacing $(y_{1}-y_{2})\equiv \Delta y\rightarrow 0$, 
$N\rightarrow \infty$
and defining $r=\Delta y.R_{1}(y)$, the scaled PCF for CSM is universal.
For $\beta=2$, it is given by
\begin{equation}
\label{corr}
R_{2}(r)=1-\frac{\sin^2(\pi r)}{(\pi r)^2}.
\end{equation}
For these systems with $\Delta y=10^{-7}$, we are in the region of the
spectrum where the particle density is high and hence interparticle 
spacing low. Here, PCF obeys Eq.(\ref{corr}). However, for  
$\Delta y=10^{-3}$ and $10^{-1}$, we are in the tail region
of the two bands. It is here that we observe a distinct deviation
from that observed in CSM. This is due to the contribution
of the long range interaction, where the third term in Eq.(\ref{Ham1})
contributes to the PCF. This becomes negligible as one makes $\Delta y$
smaller, until it disappears for $\Delta y=10^{-7}$.

\begin{figure}
\scalebox{0.44}{\includegraphics{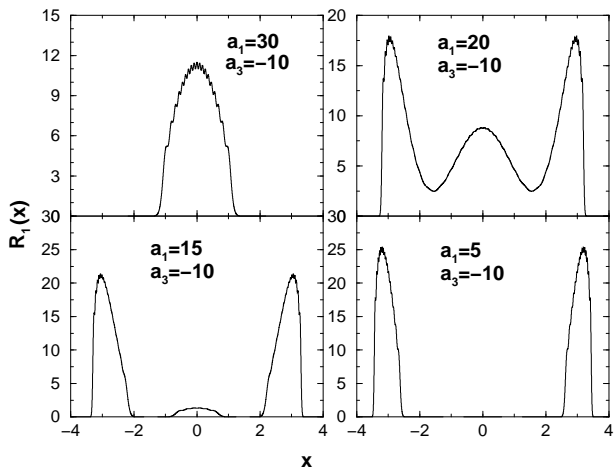}}
\caption{Particle density for the Hamiltonian given in Eq.(20), with $N=50$.
For a given $a_{3}$ and different values of the parameter $a_{1}$,
we observe a shift from single band to multiple  band structure.}
\end{figure}

For the Hamiltonian corresponding Eq.(\ref{Ham2}), it can be shown analytically that for large $N$,
that the particle density can be written in terms of the
higher moments as

\begin{eqnarray}
\nonumber
\pi R_{1}(x) &=& \gamma(-x^{10}+2a_{3}x^8-({a_{3}}^{2}+2a_{1})x^{6}\\
\nonumber
&&+(2N/\gamma+2a_{1}a_{3})x^4-({a_{1}}^{2}-2M_{2}/\gamma+2a_{3}N/\gamma)x^{2}\\
&&+(2a_{1}N+2M_{4}-2a_{3}M_{2})/\gamma)^{1/2}
\end{eqnarray}
where
$M_{p}=\int x^{p}R_{1}(x)dx$, $x$ being the scaled variable.
These moments can be calculated numerically to obtain
$R_{1}(x)$. However, a more convenient way is to directly construct
the polynomials through the recursion relation and to do the sum of
Eq.(\ref{R2}), with $x=y$. 
For $a_{3}<0$, we define 
$a_{1}\equiv a_{c}=a_{3}^{2}/4$, which corresponds to a ground state
energy $E_{0c}=-\gamma a_{c}N(1+\lambda(N-1)).$ 
 For $E_{0}<E_{0c}$, we observe a single
band structure in the particle density.
For $E_{0}> E_{0c}$, (i.e. $a_{1}< a_{c}$), the particles experience
a repulsion at the center and thus we see a transition from single band
to three band structure. As $E_{0}<<E_{0c}$, particles are completely repelled
from the center and finally collect at the two end of the spectrum, thereby
giving rise to two bands.

Thus we have studied, for both bosons and fermions, 
the ground state properties of a class of $N$-body systems
with long range  two body interaction
in addition to the inverse square potential.
We observe a distinct deviation from the ``universal''
result for the PCF, as obtained in the CSM. We also
observe that particles are arranged in bands. To study different 
thermodynamic properties, it is
necessary to study
the excitation spectrum  of such Hamiltonian.
 At finite temperature and finite
$N$, we expect  similar smoothing  of the oscillations in the
particle density \cite{gpuri} as observed in Ref.\cite{Akdeniz}.
 We will come back to this in a later
publication.

I am grateful to Dr. P. K. Panigrahi for many useful discussions.


\begin{thebibliography}{999}


\bibitem{Sutherland}
B. Sutherland, J. Math. Phys. {\bf 12}, 246 (1971).

\bibitem{Calogero}
F. Calogero, J. Math. Phys. {\bf 10}, 2191 (1969).


\bibitem{Panigrahi2}
N. Gurappa, and P. K. Panigrahi, Phys. Rev. B. {\bf 59}, 4(R) (1999). 


\bibitem{Panigrahi3}
N. Gurappa, and P. K. Panigrahi, Phys. Rev. B. {\bf 67}, 155323 (2003) and
references therein. 



\bibitem{Papenbrock1}

T. Papenbrock, Phys. Rev. A {\bf 65}, 033606 (2002).

\bibitem{Sogo}
T. Sogo, and H. Yabu, Phys. Rev. A {\bf 66}, 043611 (2002).

\bibitem{Lloyd}
Lloyd C. L. Hollenberg, and N. C. Witte, Phys. Rev. B {\bf 54}, 23 (1996).

\bibitem{Papenbrock2}
T. Papenbrock, Phys. Rev. A {\bf 67}, 041601(R) (2003).

\bibitem{Forrester}
P. J. Forrester, N. E. Frankel, T. M. Garoni, and N. S. Witte,
Phys. Rev. A {\bf 67}, 043607 (2003).


\bibitem{Panigrahi1}
R. Atre, and P. K. Panigrahi, Phys. Lett. A {\bf 317/1-2}, 46 (2003).


\bibitem{Vignolo}
P. Vignolo, A. Minguzzi, and M. P. Tosi,
 Phys. Rev. Lett. {\bf 85}, 14 (2000).




\bibitem{Beenakker}
C. W. J. Beenakker, Rev. Mod. Phys. {\bf 69}, 731 (1997). 





\bibitem{Ketterle}
W. Ketterle, and N. J. van Druten, Phys. Rev. A {\bf 54}, 656 (1996).


\bibitem{pg}
A. Pandey and S. Ghosh , Phys. Rev. Lett. {\bf 87}, 024102 (2001). 

\bibitem{gp}
S. Ghosh and A. Pandey,  Phys. Rev. E {\bf 65}, 046221 (2002). 



\bibitem{gpuri}
S. Ghosh, A. Pandey, S. Puri and R. Saha
Phys. Rev. E {\bf 67}, 025201(R) (2003).


\bibitem{Landau}
L. D. Landau and E. M. Lifshitz {\it Quantum Mechanics}
(Pergamon, New York, 1958).



\bibitem{Akdeniz}
Z. Akdeniz, P. Vignolo, A. Minguzzi and M. P. Tosi,
 Phys. Rev. A {\bf 66}, 055601 (2002).


\end{thebibliography}
\end{document}